\documentclass[a4paper]{spie}  

\usepackage{setspace}
\usepackage{tocloft}
\usepackage{lineno}
\usepackage{subcaption}
\usepackage{siunitx}
\usepackage{amsmath,amsfonts,amssymb}
\usepackage{graphicx}
\usepackage[colorlinks=true, allcolors=blue]{hyperref}

\title{Technologies and novel components for broadband splitting and coupling in pairwise and nulling interferometry}

\author[a]{Harry-Dean Kenchington Goldsmith}
\author[b]{Nemanja Jovanovic}
\author[b]{Anusha Pai Asnodkar}
\author[c]{Sanny Ahmed}
\author[d]{Elsa Huby}
\author[d]{Sylvestre Lacour}
\author[c]{Michael Fitzgerald}
\author[c]{Yoo Jung Kim}
\author[e]{Pierre Labeye}
\author[e]{Nicolas Dunoyer}
\author[f]{Michael Ireland}
\author[g]{Stephen Madden}
\affil[a]{Astrophotonics Australia, Monash, Canberra, 2904, ACT, Australia}
\affil[b]{Department of Astronomy, California Institute of Technology, 1200 East California Boulevard, Pasadena, 91125, California, United States of America}
\affil[c]{Physics \& Astronomy Department, University of California, Los Angeles, 475 Portola Plaza, Los Angeles, 90095, California, United States of America}
\affil[d]{LIRA, Observatoire de Paris, Université PSL, CNRS, Sorbonne Université, Université de Paris, 5 place Jules Janssen, Meudon, France}
\affil[e]{Commissariat à l’Energie Atomique, Leti, 17 rue des Martyrs, 38054, Grenoble, France}
\affil[f]{Astralis-AITC, Research School of Astronomy and Astrophysics, Australian National University, Mount Stromlo Observatory, Cotter Road, Canberra, 2611, ACT, Australia}
\affil[g]{Department of Quantum Science, Research School of Physics, Australian National University, 60 Mills Rd, Canberra, 2601, ACT,  Australia}
\authorinfo{Further author information: (Send correspondence to H-D.K.G.)\\H-D.K.G.: E-mail: harrykenchington@astrophotonics.com.au }

\begin{document} 
\maketitle
\begin{abstract}

Passive and active photonic components are central to the continued development of astronomical photonic integrated circuits~(PICs), although achieving broadband achromatic performance remains a significant challenge.

This work details novel, broadband, evanescent tri-couplers and directional-couplers, and a proposed achromatic intensity modulator with active control, in the astronomical J- and H-bands for photonics in interferometry. Silicon nitride and silicon oxide remain key materials for low-loss, broadband PICs in this region. Customised designs based on standard components, such as tapered directional couplers and tapered tri-couplers, can provide excellent broadband performance across the 0.95–1.8\,\si{\um} range.

The silicon nitride technology from STMicroelectronics was investigated for use in a pairwise-combination chip in the astronomical J-band, with upgraded splitters and couplers simulated to replace the standard components. Tapered tri-couplers and tapered directional couplers were simulated to determine the broadband response over the waveband. Optimised tapered tri-couplers achieve $<1$\% excess loss across the J-band. Optimised tapered directional couplers provide broadband 50:50 and 40:60 splitting suitable for beam combination.

The silicon oxide technology from Enablence uses a germanium-doped silicon oxide layer as a waveguide core, clad in undoped silicon oxide to form a relatively low-confinement waveguide in the astronomical H-band. This work focuses on the nulling interferometer using a two-dimensional tri-coupler. By default, the tri-coupler will facilitate destructive interference in the central output port when injected with two equal-intensity electric fields at a 180° phase offset. This is how to null star-light in a nulling interferometer. Maximal exoplanet throughput occurs at 0° phase offset. Optimised tapered tri-couplers achieve less than 2.2\% exoplanet throughput loss across the H-band while retaining broadband phase sensing capability. These tapered tri-couplers were also shown to be capable of fringe tracking.

A chromatically controlled achromatic intensity modulator (CCAIM) is also introduced. The device combines a tapered directional coupler with a thermo-optic phase shifter to provide broadband programmable attenuation, enabling active intensity balancing for nulling interferometers.

Future work will extend these concepts to mid-infrared chalcogenide platforms for astronomical nulling interferometry. This is the regime that will best serve space missions for detecting Earth-like exoplanets. At longer wavelengths, where silicon becomes absorptive, chalcogenide glasses offer a promising alternative by balancing high optical confinement with efficient fibre coupling.

Together, these developments illustrate the growing maturity of broadband photonic components for astronomical interferometry and nulling. Continued progress in materials, fabrication precision, and device design will be essential for enabling fully integrated, low-loss, and highly scalable photonic interferometers across the near- and mid-infrared spectrum.

\end{abstract}


\section{Introduction}
\label{sec:intro}  

Photonic integrated circuits~(PICs) are emerging as a key technology for future ground- and space-based astronomical instruments, enabling smaller, lighter, and more environmentally robust systems. PICs provide a platform for integrating adaptive optics, beam combination, spectroscopy, and single-photon detection onto a single monolithic chip.

Astronomical interferometry dates back to Michelson's pioneering measurements of Jupiter's moons~\cite{Michelson1891}. Today, interferometric photonic beam combiners are used at the Very Large Telescope Interferometer~(VLTI) through the GRAVITY instrument~\cite{Benisty2009} and in aperture-masking interferometry with the Fibred Imager for a Single Telescope~(FIRST) at the Subaru Telescope~\cite{Huby2012,Perrin2006}.

These instruments rely on carefully engineered splitters and combiners for pairwise interference, allowing complete reconstruction of the incoming wavefront. Nulling interferometers further exploit the spatial separation between a star and an orbiting exoplanet while transmitting light from an off-axis exoplanet located near the first transmission maximum at $\lambda/2B$ ($B$: the interferometric baseline).

Several major instruments already employ nulling interferometers on-chip, including the Guided Light Interferometric Nulling Technologies~(GLINT)~\cite{Norris2014} at the Subaru Telescope and the Nulling Observations of exoplaneTs and dusT~(NOTT)~\cite{Sanny2026} as part of the Asgard Instrument Suite~\cite{Defrre2024} at the VLTI.

Selecting the right material for a project depends on the technology available for the wavelength range. Silicon photonics dominates from the visible to the near-infrared. Silicon nitride~(Si\textsubscript{3}N\textsubscript{4}) and silicon oxide~(SiO\textsubscript{2}) have low material losses in the H- and J-astronomical wavebands. Customised designs based on standard components, such as tapered directional couplers, tapered tri-couplers, and MMIs with optimised input tapers, have been shown in SiO\textsubscript{2} to provide excellent broadband performance across the 1.5–1.8\,\si{\um} range~\cite{KenchingtonGoldsmith2026}.

At longer wavelengths, where exoplanets glow brightly~\cite{Marois2008}, germanium-silicon platforms are options for high-transmission astrophotonics~\cite{Turpaud2024}; however, when Si becomes absorptive, chalcogenide glasses~(ChG) offer a powerful alternative owing to their broad transparency window and the ability to tailor the refractive index through glass composition. A combination of sulphur and selenium glasses, for example, has demonstrated low propagation loss in the L-band~\cite{Ma2013}, and broadband MMIs fabricated in these materials have also demonstrated nulling interferometry in the L-band~\cite{KenchingtonGoldsmith2016,KenchingtonGoldsmith2017b}.

Indium Gallium Arsenide~(InGaAs) and Indium Phosphide~(InP) also have low absorption in the MIR, with recent technology creating a sub-1\,dB/cm waveguide from 5 to 11\,\si{\um} in wavelength~\cite{Montesinos-Ballester2024}. To date, a complete library of broadband beam splitters and combiners has not yet been demonstrated for this platform, so its suitability for astronomical interferometry remains uncertain. As with ChG, separate PIC designs may ultimately be required for different wavebands to span the entire L-, M-, and N-band. These materials could produce PICs compatible with space telescopes such as the Habitable Worlds Observatory (HWO)~\cite{feinberg2026} and the Large Interferometer For Exoplanets~(LIFE) project~\cite{Quanz2022} that could detect ozone on an Earth-like exoplanet (at 9.7\,\si{\um})--an indicator of life~\cite{Segura2003}.

This work presents recent developments on next-generation splitters and couplers designed for the PLANETES project in the J-band in Sec.\,\ref{sec:PLANETES}, and for a nulling interferometer for Caltech's astrophotonics team in Sec.\,\ref{sec:Caltech}. The focus is on broadband passive photonic components that improve beam splitting and beam combination in both pairwise and nulling architectures. A new device, an achromatic intensity modulator, is presented in Sec.\,\ref{sec:Intensity_matching} to actively balance optical intensity between the interferometer arms and thereby improve the achievable null depth. Finally, a roadmap towards mid-infrared astrophotonic platforms for L- and N-band astronomical interferometry is presented.

\section{Silicon Nitride broadband components}
\label{sec:PLANETES}

Silicon Nitride (Si\textsubscript{3}N\textsubscript{4}), clad with Silicon Oxide (SiO\textsubscript{2}), is a well-established photonic platform and has recently been evaluated for an on-chip spectrometer~\cite{Gatkine2024}, alongside the doped silicon oxide platform discussed in Sec.\,\ref{sec:Caltech}. The principal advantage of Si\textsubscript{3}N\textsubscript{4} is its higher index contrast, which enables small bend radii with low propagation loss, as well as compact couplers and splitters. This facilitates the implementation of multi-telescope beam combiners and more complex photonic circuits within a reduced footprint. A drawback of the platform is the increased coupling loss associated with the tightly confined (often sub-wavelength) fundamental mode. Addressing this challenge is left to the manufacturers but is discussed somewhat in Sec.\,\ref{sec:MIR}.

The GRAVITY PIC~\cite{Gillessen2010} at the VLTI and the Fibred Imager for a Single Telescope~(FIRST) PIC at the Subaru Telescope~\cite{Barjot2020} are first-generation instruments utilising Y-junctions and directional couplers.

An equivalent chip design is being developed within the PLANETES project~\cite{Sarrazin2026}, which aims to provide direct imaging capabilities for exoplanet detection and characterisation at the VLTI in the astronomical J-band (specifically from 1080 to 1350\,nm), expanded to span 950 to 1350\,nm.

This four-telescope instrument is conceptually similar to the GRAVITY PIC but operates in the J-band rather than the astronomical K-band. It employs the same fundamental building blocks as earlier PICs; however, instead of conventional pairwise interferometry, it uses ABCD couplers that encode the full complex wavefront in a single measurement~\cite{Benisty2009}. The resulting signals can then be reconstructed using a visibility-to-pixel matrix~(V2PM)~\cite{Lacour2014}.

Included in the first fabrication run were experimental second-generation components intended to increase throughput, including tapered tri-couplers for lossless splitting ratios and tapered directional couplers designed to achieve equal-power coupling of telescope inputs.

The simulations presented here were performed using RSoft BeamPROP and considered only the TE polarisation. At a wavelength of 1100\,nm, the refractive indices of Si\textsubscript{3}N\textsubscript{4} (2.01) and SiO\textsubscript{2} (1.45) give an index contrast of 0.56. Owing to the high optical confinement, waveguides as narrow as 600\,nm and 400\,nm high were employed and fabricated by STMicroelectronics~\cite{Cremer2026}. As noted above, the mode field diameter is approximately 1,\si{\um}, substantially smaller than the typical fibre mode-field diameter of 6.6,\si{\um}~\cite{Thorlabs}, which is approximately 5.5\% throughput based on the overlap function.

Addressing this coupling challenge is beyond the scope of the present work, but it remains an important consideration for future development of the platform. This work focuses on the second-generation splitters and couplers.

\subsection{Tapered tri-coupler splitters}
\label{sec:TTC}

A standard tri-coupler can be designed as a replacement for a Y-junction using the tri-coupler beat length, $L_{\pi}$,

\begin{equation}
L_{\pi}=\frac{\lambda}{n_A-n_B},
\label{eq:BeatLength}
\end{equation}

where $n_A$ and $n_B$ are the effective indices of the two symmetric supermodes of the tri-coupler (excluding the antisymmetric mode), and $\lambda$ is the wavelength~\cite{Donnelly1986}.

For a two-dimensional platform, unlike the three-dimensional tri-couplers reported by Martinod~et~al.~\cite{Martinod:21}, this relationship is only valid for excitation of the central input waveguide. Achieving achromatic splitting from the central waveguide to the two outer waveguides in a planar tri-coupler requires tapering of the outer waveguides~\cite{KenchingtonGoldsmith2026}.

\begin{figure}[ht] 
    \centering 
        \includegraphics[angle=-90, width=0.45\linewidth]{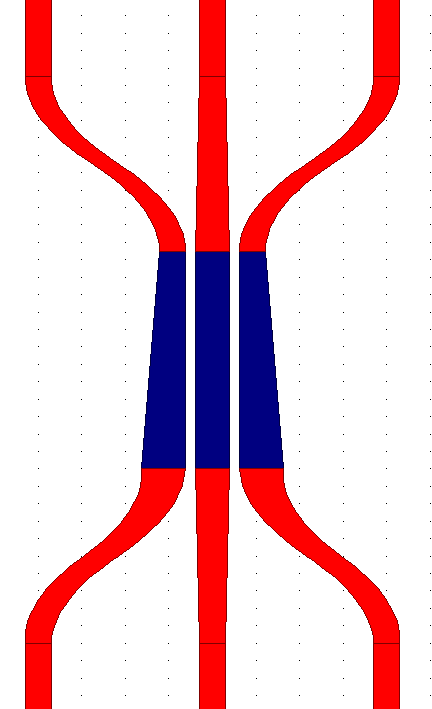} 
    \caption{Tapered tri-coupler CAD.} 
    \label{fig:Tapered_Tricoupler_CAD} 
\end{figure}

Figure~\ref{fig:Tapered_Tricoupler_CAD} shows the RSoft CAD model used to simulate the tapered tri-coupler. The coupling-region, highlighted in blue, consists of a constant-width central waveguide and two outer waveguides whose widths vary along the interaction region while maintaining a constant waveguide gap of 220\,nm. To reduce the parameter space, the outer waveguide widths are described by a symmetric deviation from the central waveguide width, such that $W_{L,S}=W_{C}\pm\Delta W$. 

The central waveguide may be widened before entering the coupling-region if required, but its width remains constant throughout the interaction region. Tapered transitions connect the coupling-region to the global waveguide width of 600\,nm.

The bend radius is approximately 500,\si{\um} (200,\si{\um} end-to-end and 4,\si{\um} lateral displacement), resulting in a simulated 1\% loss. Increasing the bend radius alters the effective interaction region, which begins shortly before and ends shortly after the coupling region. The bends used in this work were the default RSoft CAD S-bends. For the tapered tri-coupler, neither the bend profile nor the bend radius significantly affected the splitting ratio and were therefore not investigated further.


Any optical power emerging from the central output port is treated as splitter loss. To minimise this loss, the coupling-region and the widths of the three waveguides were optimised to reduce the power in the central output. The waveguide gap was held at a constant 200\,nm throughout the coupling-region and was not included as an optimisation parameter. The central waveguide width was fixed at 800\,nm to simplify the optimisation and to permit larger values of $\Delta W$ without producing waveguides that were too narrow to support guided modes.

\begin{figure}[ht]
    \centering
    \includegraphics[width=0.7\linewidth]{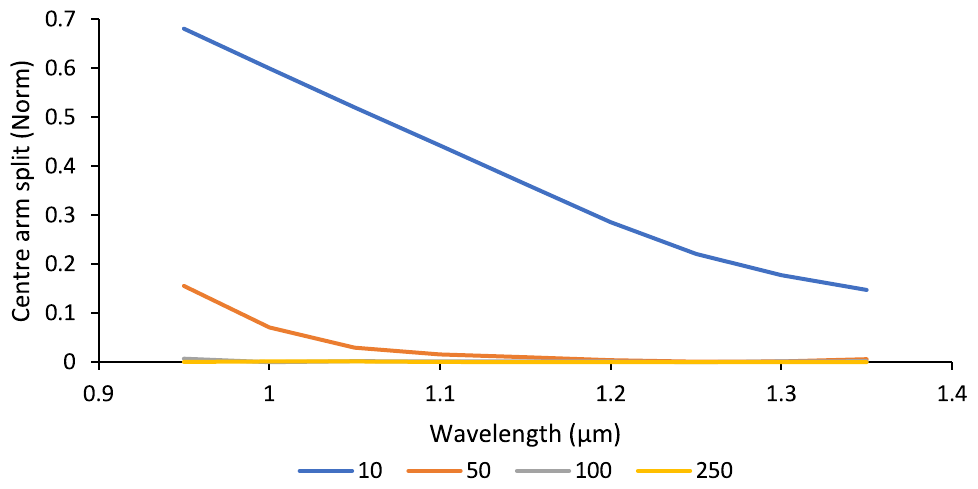}
    \caption{The simulated centre light intensity in the tri-coupler for different coupling-regions.}
    \label{fig:TapaeredTrioupler_Length_Loss}
\end{figure}

To explore the length parameter space, the $\Delta W$ was set to 200\,nm. The length simulation is shown in Fig.\,\ref{fig:TapaeredTrioupler_Length_Loss}. These lengths range from so small that the tapered tri-coupler behaves similarly to a standard tri-coupler to large enough that centre light is mitigated significantly over the entire J-band. 

It shows that a 50\,\si{\um} coupling-region has a loss of 15\% at shorter wavelengths, and increasing the coupling-region to 100 or 250\,\si{\um} decreases the loss to $<1\%$. It highlights the adiabatic response of the tri-coupler, with the loss decreasing as the length increases, with the ideal response occurring at a relatively small length compared to the component lengths in Sec.\,\ref{sec:Caltech}.

For the small portion of light remaining in the central waveguide not to interfere with the rest of the chip, the light must be directed away from the end PIC facet or else risk coherent light scattered into the PIC mixing with the guided light. Also note that due to fabrication limitations~\cite{KenchingtonGoldsmith2024}, some loss is always expected, and redirection is recommended to displace this light.

\begin{figure}[ht]
    \centering
    \includegraphics[width=0.7\linewidth]{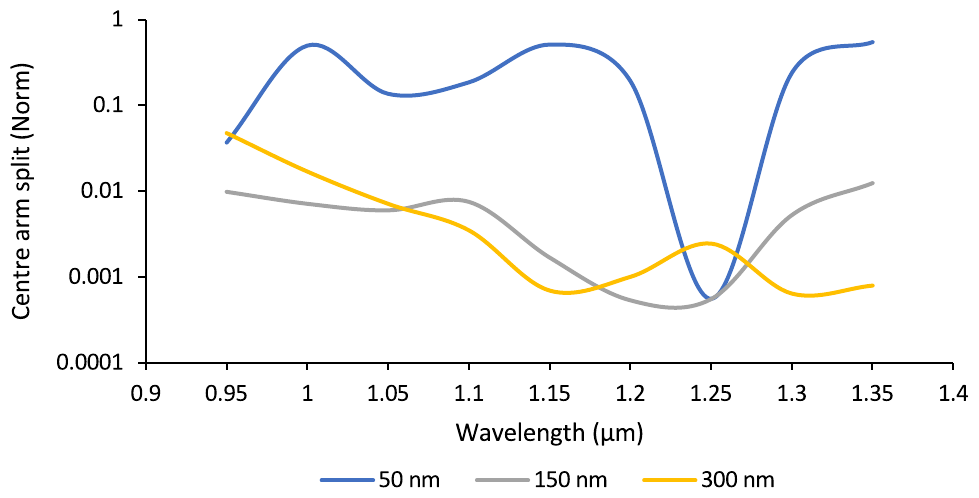}
    \caption{The simulated centre light intensity in the tri-coupler for different $\Delta W$.}
    \label{fig:TapaeredTrioupler_Width_Loss}
\end{figure}

The coupling-region length was fixed at 250,\si{\um} to investigate the dependence on $\Delta W$. This provides a sufficiently long interaction region, noting that increased adiabaticity is generally associated with improved achromatic performance~\cite{Hsiao2010}. Figure~\ref{fig:TapaeredTrioupler_Width_Loss} shows the effect of varying $\Delta W$ on the power emerging from the central output port.

A small value of $\Delta W$ produces a narrow low-loss operating region characteristic of a conventional tri-coupler, whereas larger values provide significantly improved broadband performance. The minimum simulated loss was below 1\% for $\Delta W=300$\,nm across the entire J-band. Further optimisation of the interaction length could reduce this residual loss; however, these simulations demonstrate that the tapered tri-coupler can split light equally between the two outer waveguides while maintaining less than approximately 1\% power in the central output.

\begin{figure}[ht]
    \centering
    \includegraphics[width=0.7\linewidth]{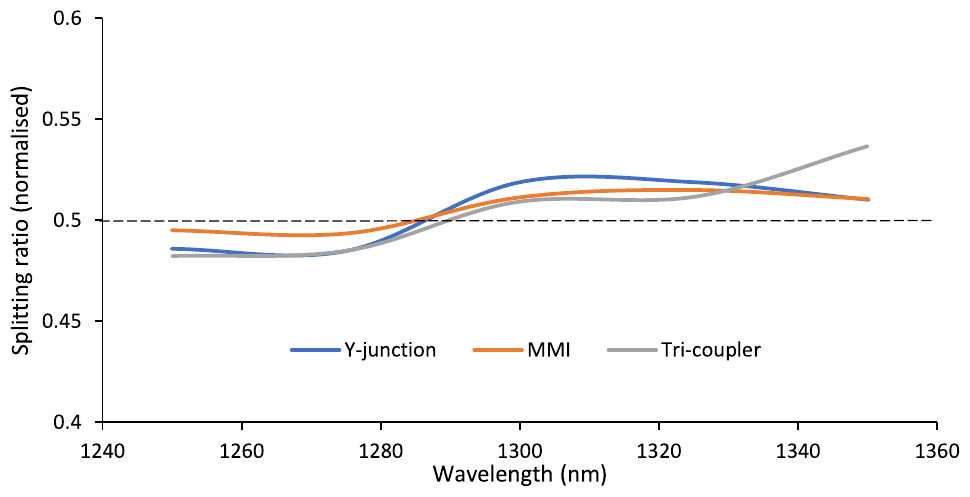}
    \caption{Comparative splitting ratio for the Y-junction, 1x2~MMI, and the tapered tri-coupler.}
    \label{fig:TaperedTricoupler_Measured}
\end{figure}

Preliminary measurements of the tapered tri-coupler, together with two comparison splitters--a Y-junction and a 1×2 MMI--were performed by colleagues in the Laboratoire d'électronique des technologies de l'information (LETI) over a limited wavelength range. The results are shown in Fig.\,\ref{fig:TaperedTricoupler_Measured}.


The measured splitting ratios are not exactly 50:50 across the entire wavelength range for any of the devices. In an ideal symmetric structure, a 50:50 split would be expected; therefore, the observed deviations indicate some degree of fabrication asymmetry, consistent with previous experimental results reported for tapered tri-couplers~\cite{KenchingtonGoldsmith2024}. The supplier also reported negligible excess loss in the tapered tri-coupler. Independent measurements will be required to determine the amount of optical power emerging from the central output port.


These preliminary results indicate that the tapered tri-coupler is a viable alternative to a Y-junction in applications where Y-junctions and MMIs exhibit significant excess loss. Future work should compare the throughput of all three splitter types and extend the measurements across the full J-band to better quantify their relative performance and limitations.

\pagebreak
\subsection{Tapered directional couplers}
\label{sec:TDC}

\begin{figure}[ht]
    \centering
    \includegraphics[width=0.5\linewidth]{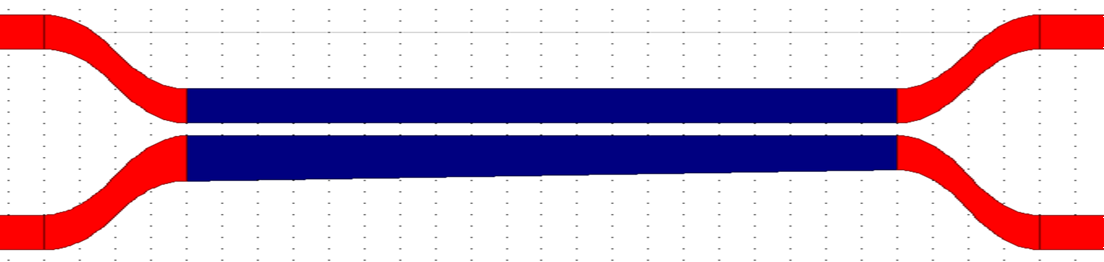}
    \caption{CAD tapered directional coupler.}
    \label{fig:TaperedDC}
\end{figure}

The tapered directional coupler operates as a broadband 2×2~splitter used as a beam combining component in a multi-telescope pairwise interferometry PIC. The figure of merit for these simulations was whether the split is close to 50\%: a lower range of 40:60\% was selected for this work. The device evaluated in this work is shown in Fig.\,\ref{fig:TaperedDC}. It consists of a constant-width 600\,nm waveguide and a tapered waveguide, with both waveguides returning to the same width at the end of the coupling-region (in blue).

\begin{figure}[ht]
    \centering
    \begin{subfigure}{0.7\textwidth}
            \includegraphics[width=0.9\linewidth]{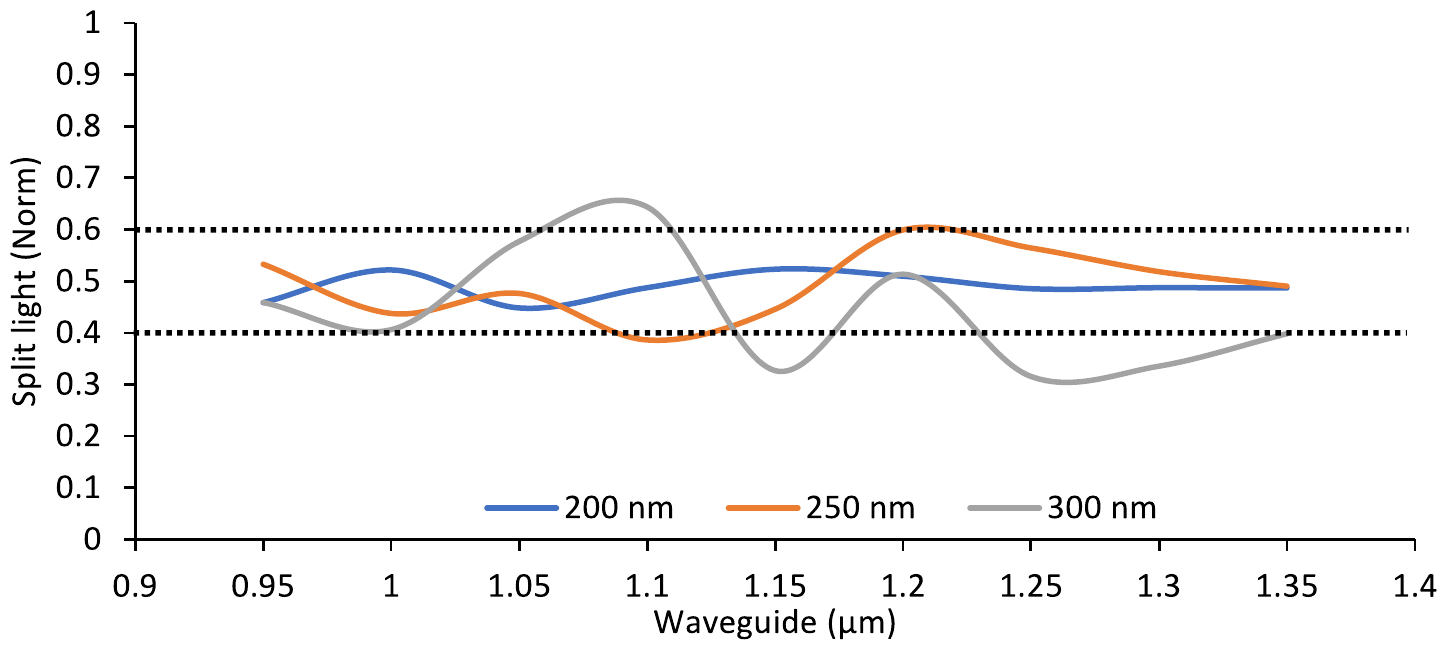}
        \caption{$<600$\,nm}
                \label{fig:TaperedDC_Width_Small}
    \end{subfigure}
    \begin{subfigure}{0.7\textwidth}
            \includegraphics[width=0.9\linewidth]{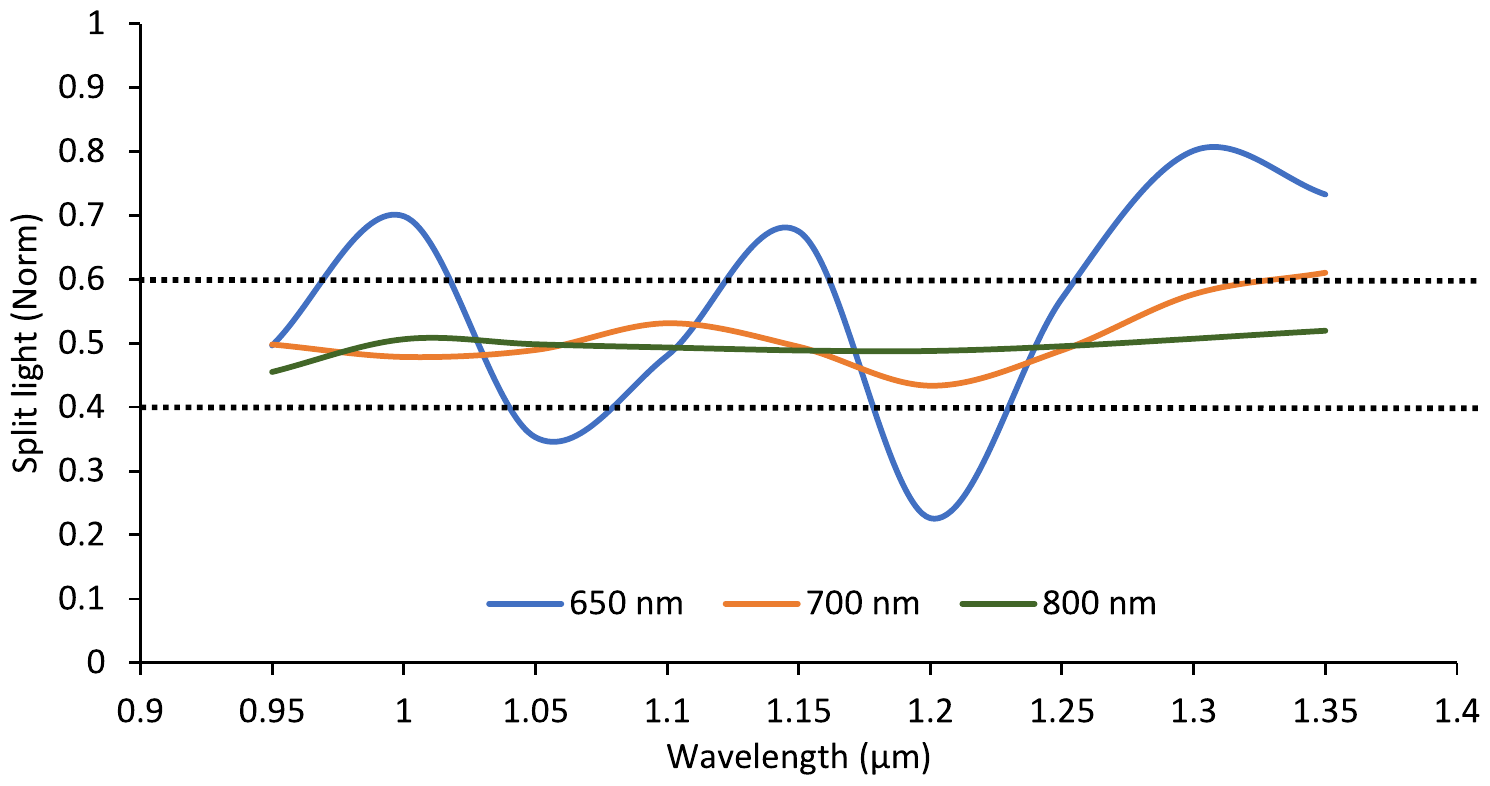}
        \caption{$>600$\,nm}
        \label{fig:TaperedDC_Width_large}
    \end{subfigure}
    \caption{The splitting ratio for tapered directional couplers for waveguides widths (a) $<600$\,nm and (b) $>600$\,nm.}
    \label{fig:TaperedDC_Width}
\end{figure}

As with the tapered tri-coupler, only the waveguide widths and coupling-region length were varied in this study, while the waveguide gap was held constant throughout the coupling region. The optimisation for an even splitting ratio was separated into two groups according to whether the tapered-waveguide input width is smaller or larger than the global waveguide width of 600\,nm. The resulting splitting ratios are shown in Fig.\,\ref{fig:TaperedDC_Width}.

The smaller-width tapered directional couplers, shown in Fig.\,\ref{fig:TaperedDC_Width_Small}, exhibit low-amplitude oscillations about the 50\% splitting ratio and remain within the 40:60\% range across the full bandwidth for an input width of 200\,nm. Larger widths deviate further from the desired equal split. However, reducing the waveguide width also increases the mode size and decreases optical confinement. This may reduce the reliability of beam-combiner simulations because the optical mode becomes weakly confined.

The larger-width tapered directional couplers, shown in Fig.\,\ref{fig:TaperedDC_Width_large}, maintain a splitting ratio within the 40:60\% range across the full bandwidth for an input width of 800\,nm. For a width of 650\,nm, the response exhibits a periodic wavelength dependence characteristic of a conventional directional coupler, as expected when the taper approaches the global waveguide width. Larger waveguide widths do not suffer from weak confinement but may become multimodal at shorter wavelengths. No evidence of multimodal behaviour is apparent in these simulations; however, a full modal analysis is required to confirm the suitability of these wider waveguides.

\begin{figure}[ht]
    \centering
    \includegraphics[width=0.65\linewidth]{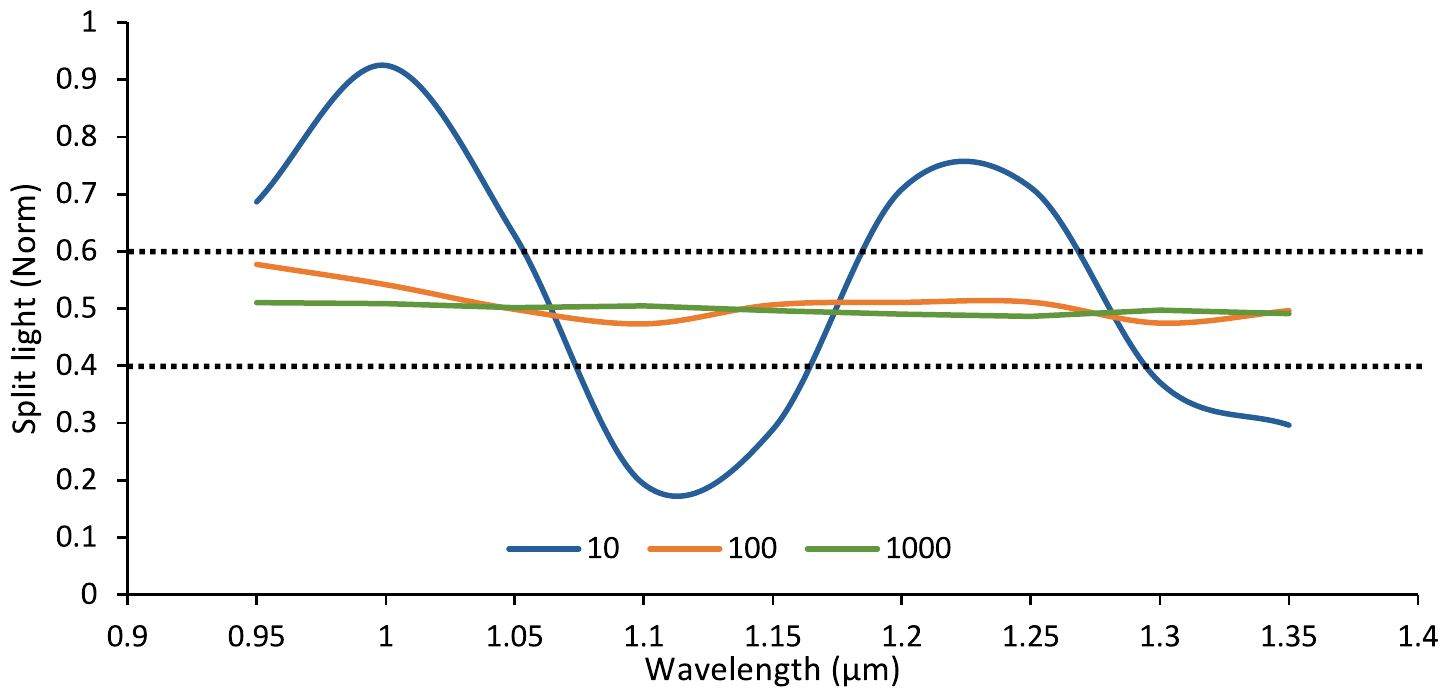}
    \caption{Splitting ratio of the tapered directional coupler for different coupling region lengths.}
    \label{fig:TaperedDC_Length}
\end{figure}

The results shown in Fig.\,\ref{fig:TaperedDC_Width} indicate that waveguide widths further from the global width of 600\,nm produce flatter broadband responses. This behaviour suggests that adiabaticity alone does not determine performance in these structures and that additional mode-evolution effects contribute to the observed coupling characteristics. Figure\,\ref{fig:TaperedDC_Length} shows the splitting performance of an 800\,nm tapered directional coupler for various coupling-region lengths.

Figure\,\ref{fig:TaperedDC_Length} shows that a short coupling-region produces the periodic response characteristic of a conventional directional coupler. As the coupling-region increases, the response becomes progressively flatter and more achromatic, consistent with the expected behaviour of an adiabatic coupler.

Tapered beam combiners were previously investigated in the latest FIRST PIC update~\cite{KenchingtonGoldsmith2024}, which highlighted fabrication challenges associated with broadband TE and TM evanescent couplers operating in the visible spectrum. The present work considers only TE polarisation; however, experimental verification will be required to determine whether the fabricated devices maintain a splitting of $50\pm10$\% or better across the J-band. These couplers have not yet been measured, although the deviations from ideal splitting observed in Fig.\,\ref{fig:TaperedTricoupler_Measured} suggest that some departure from the simulated performance should be expected.

These components represent potential upgrades to the baseline PLANETES beam-combiner architecture, providing improved splitting and coupling performance across the J-band.

\section{Silicon oxide broadband components}
\label{sec:Caltech}

SiO\textsubscript{2} is an alternative technology for broadband couplers and splitters. This section focuses on work undertaken for the California Institute of Technology to design a broadband nulling interferometer using a germanium-doped core SiO\textsubscript{2} waveguide clad in an undoped SiO\textsubscript{2} ($n=1.444$, $k=0$) material. This material has a comparatively low refractive index contrast:  2\% compared to 39\% for the Si\textsubscript{3}N\textsubscript{2} material. The wavelength range for this work was also shifted to the astronomical H-band, 1500 to 1800\,nm, and is built on the arrayed waveguide gratings developed by Gatkine~et~al~\cite{Gatkine2024} that tested the SiO\textsubscript{2} platforms from Enabelance and a Si\textsubscript{3}N\textsubscript{2} from LioniX. 

This SiO\textsubscript{2} platform was selected because it provides square waveguides, 3.4$\times$3.4\,\si{\um}, for zero birefringence by default. Like with the Si\textsubscript{3}N\textsubscript{2} simulations in Sec.\,\ref{sec:PLANETES}, these are limited only to TE light for simplicity, but beginning with a square waveguide will hopefully make adapting these components for accepting TE and TM light simultaneously easier. The bend radius for this work was set to 850\,\si{\um}, which produces approximately 3.5\% loss (at a 1800\,nm wavelength) and should be increased in future work, but was selected to maintain a compact circuit. 

This work was motivated in part by photonic lantern~(PL) architectures that combine a mode-selective PL with a nulling interferometer PIC. In these systems, the PL's fundamental mode contains most of the stellar light, while the higher-order modes contain varying amounts of residual stellar leakage~\cite{Xin2022}. A nulling interferometer, Sec.\,\ref{sec:Nulling}, suppresses the coherent stellar light, but achieving a deep null requires active intensity matching between the fundamental mode and the residual leakage channels. This challenge motivates the chromatically controlled achromatic intensity modulator presented in Sec.\,\ref{sec:Intensity_matching}.

\subsection{Nulling interferometry}
\label{sec:Nulling}

Nulling interferometry suppresses on-axis starlight through destructive interference while allowing off-axis planetary light to propagate through the instrument. Nulling interferometry in photonics has been achieved in the mid-infrared using chalcogenide glass~(ChG) with MMIs~\cite{KenchingtonGoldsmith2017a} and directional couplers~\cite{Gretzinger2019b}, with the latter forming the basis of the NOTT instrument~\cite{Gretzinger2019a, Sanny2026} at the VLTI that employs three-dimensional tri-couplers. A related approach is employed in the GLINT PIC~\cite{Martinod:21} to achromatically null light in the astronomical H-band.

The equidistant three-dimensional tri-coupler is a promising candidate for such an architecture since it nulls achromatically when asymmetric light (equal intensity with a 180° phase offset) is injected into two waveguides. For a two-dimensional tapered tri-coupler, as displayed in Sec.\,\ref{sec:TTC}, the same achromatic nulling occurs when the outer waveguides are injected with asymmetric light. This is how the starlight is to be nulled. For the exoplanet light, consider that the Si\textsubscript{3}N\textsubscript{2} tapered tri-coupler was shown to be an achromatic splitter, indicating, through symmetry, that light will be directed into the central nulled waveguide when symmetric light is input into the outer waveguides. Hence, when measuring the tri-coupler as a splitter, light emerging from the central output port corresponds to exoplanet light lost into the photometric channel and is therefore used as the optimisation metric.

Three nulling interferometer tri-couplers were previously simulated in Kenchington~Goldsmith~et~al.~\cite{KenchingtonGoldsmith2026}. That work compared a conventional directional coupler, an optimised MMI, and a tapered tri-coupler. We include here a further optimisation of the tapered tri-coupler with additional detail regarding the methodology used to select the required waveguide widths for producing less exoplanet light loss over the full H-band.



The same parameters are optimised for the tapered tri-coupler as the Si\textsubscript{3}N\textsubscript{2} work in Sec.\,\ref{sec:TTC} but for the SiO\textsubscript{2} platform: The outer widths ($W_S$ and $W_L$) were the only parameters varied for this tri-coupler using a central waveguide width fixed at 3.4\,\si{\um}. The waveguide gap was fixed at the minimum fabricable value of 2\,\si{\um}. The coupling-region length was previously one of the key parameters in the tapered tri-coupler's splitting optimization, and for the larger parameter scan it was here too. For brevity, only the last length was used: 580\,\si{\um}. 

The differences in width and length from those reported in the Si\textsubscript{3}N\textsubscript{2} work are indicative of the step index difference between the platforms. The SiO\textsubscript{2} waveguide widths are approximately six times larger than those used in the Si\textsubscript{3}N\textsubscript{2} platform, but the length is closer to three times larger. Although it was not expected to be a one-to-one recreation, it may mean that the adiabatic property of the tapered tri-coupler is not applicable; instead, an achromatic response must be located within the parameter space. The smaller length was selected to align with the length of the other simulated nulling interferometers~\cite{KenchingtonGoldsmith2026}.

\begin{figure}[ht]
    \centering
    \begin{subfigure}{0.32\textwidth}
            \includegraphics[width=0.9\linewidth]{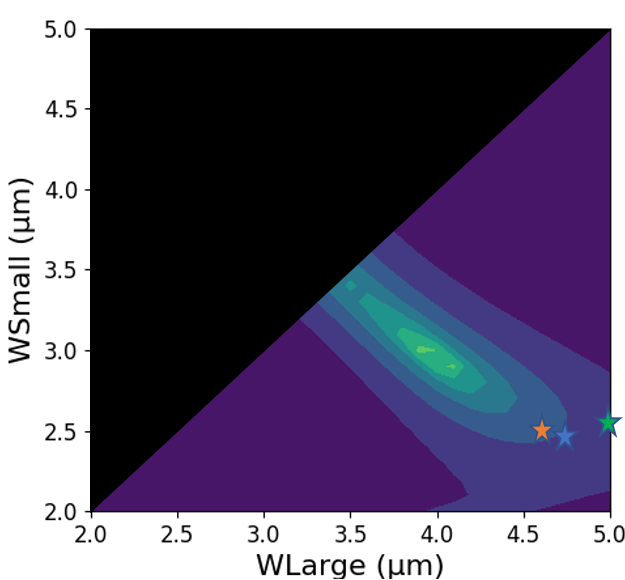}
    \caption{1400\,nm}
        \label{fig:TaperedTriCoupler_Width1P4}
    \end{subfigure}
        \begin{subfigure}{0.32\textwidth}
            \includegraphics[width=0.9\linewidth]{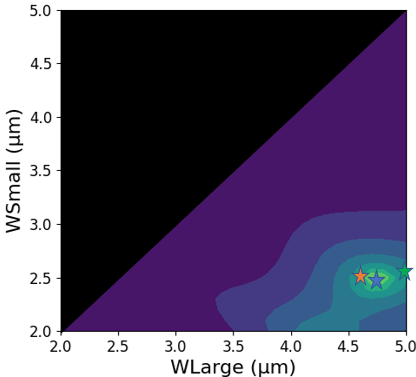}
    \caption{1600\,nm}
        \label{fig:TaperedTriCoupler_Width1P6}
    \end{subfigure}
        \begin{subfigure}{0.32\textwidth}
            \includegraphics[width=1.07\linewidth]{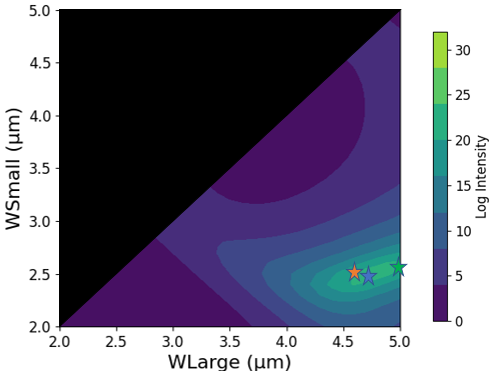}
    \caption{1800\,nm}
        \label{fig:TaperedTriCoupler_Width1P8}
    \end{subfigure}
    \caption{The central waveguide power (loss), shown on a decibel scale, for tapered tri-couplers at various $W_S$ and $W_L$ combinations, for the (a) 1400\,nm (b) 1600\,nm, and (c) 1800\,nm wavelengths, where the lighter (green) areas have the least light. The stars highlight the selected parameters (colours matching those in later figures).}
    \label{fig:TaperedTriCoupler_WidthWavelength}
\end{figure}

To determine the best tapered tri-coupler parameters, the central port light was simulated for $W_S$ and $W_L$ between 2 and 5\,\si{\um}, which are the limits determined to be within fabrication limits and ensuring single-mode operation, respectively, for three equidistant wavelengths over the waveband. These are shown in Fig.\,\ref{fig:TaperedTriCoupler_WidthWavelength}, the cases for $W_S>W_L$ blacked out. Each image highlights areas in which the exoplanet throughput is highest, but not necessarily for all wavelengths.

This methodology was inspired by the multi-segmented tri-coupler by Li~et~al.~\cite{Li2025}. They simulated the central output port for the tapered tri-coupler for different wavelengths and found low-loss regions in the parameter space, each normalised to the total output. Increasing the number of optimisation wavelengths improves the robustness of the solution; however, three wavelengths were sufficient for the present study.

The parameters selected in this work are highlighted with the star shapes in Fig.\,\ref{fig:TaperedTriCoupler_WidthWavelength} and included in Table\,\ref{tab:WidthsSelection}. 

\begin{table}[ht]
    \centering
    \caption{The selected widths for the simulated tapered tri-couplers and compared to their star colours in Fig.\,\ref{fig:TaperedTriCoupler_WidthWavelength} and those below.}
    \begin{tabular}{|c|c|c|c|}\hline
    No. & Colour & $W_S$\,\si{\um} & $W_L$\,\si{\um} \\\hline
    1 & Blue    & 2.5   & 4.7  \\
    2 & Orange  & 2.55  & 4.6  \\
    3 & Green   & 2.6  & 5.0    \\\hline
    \end{tabular}
    \label{tab:WidthsSelection}
\end{table}

Each begins at $W_S=2.55\pm0.05$\,\si{\um} with the $W_L$ varying from 4.6 to 5\,\si{\um}. These were selected to show the impact of a small shift in waveguide width (indicative of common fabrication errors) from the optimal exoplanet throughput selection (No. 1), and to illustrate the trade-off between exoplanet throughput and sensing capability--the ability to estimate the phase imbalance through monitoring the tri-coupler's photometric channels and correct it to drive the null deeper--otherwise known as fringe tracking~\cite{Klinner-Teo2022}.  

\begin{figure}[ht]
    \centering
    \includegraphics[width=0.60\linewidth]{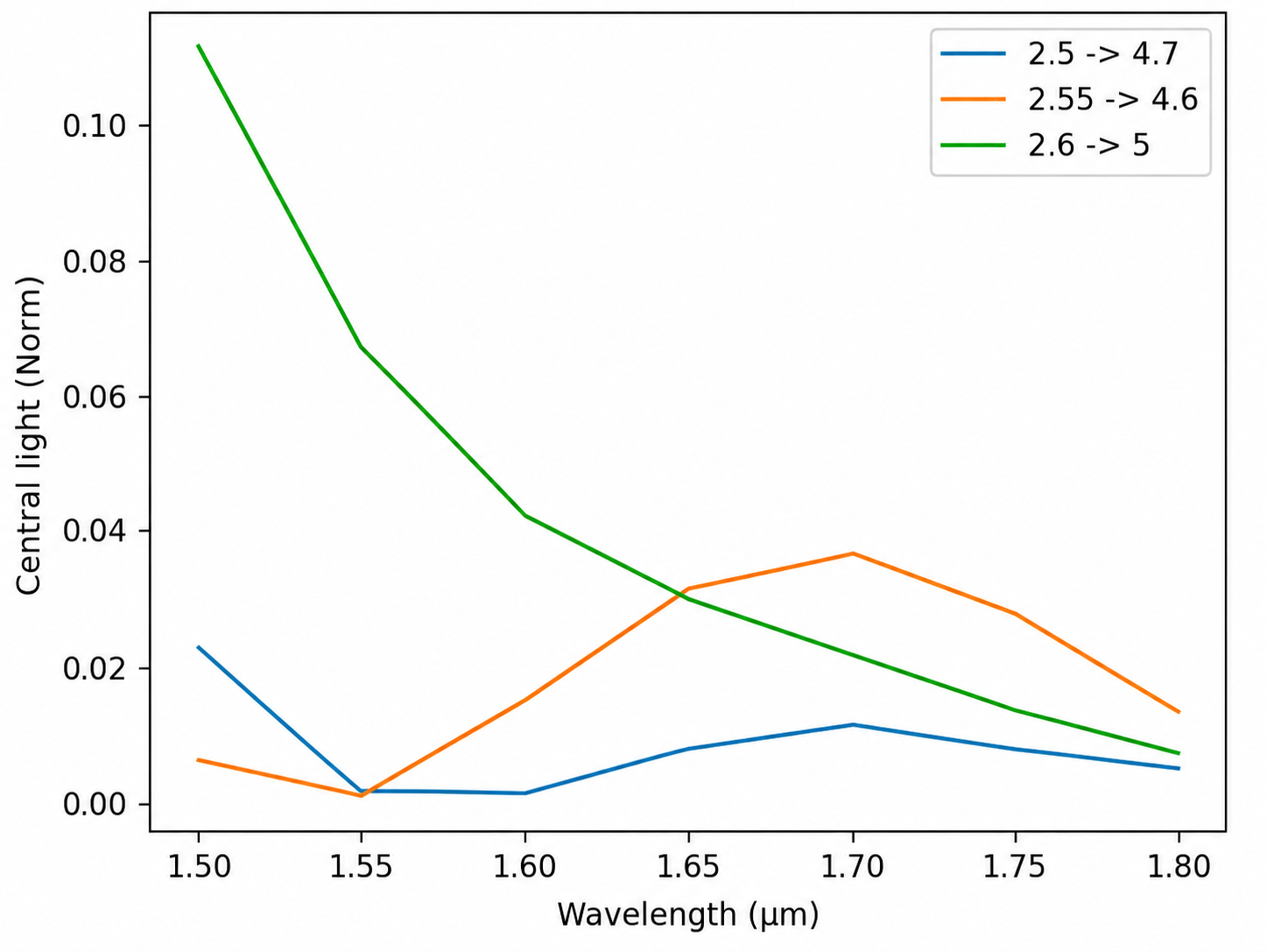}
    \caption{The simulated light remaining in the tapered tri-coupler's central output port, equivalent to exoplanet light lost when used as a nulling interferometer.}
    \label{fig:TaperedTriCoupler_ExThrough}
\end{figure}

The central port light intensity over the entire waveband is shown in Fig.\,\ref{fig:TaperedTriCoupler_ExThrough} for each parameter combination in Table\,\ref{tab:WidthsSelection}. 

Figure\,\ref{fig:TaperedTriCoupler_ExThrough} shows that a flat exoplanet loss with wavelength is achievable. The best parameters are $W_S=2.5\text{ and }W_L=4.7$\,\si{\um} with a $<2.2$\% exoplanet light loss over the waveband. It also shows that a slight deviation in the waveguide widths (the orange curve relative to the blue) increases the loss to $4.5$\% toward the long-wavelength end of the waveband, indicating significant sensitivity to fabrication-induced width variations. This does not include errors in the central arm or an asymmetry between the output arms, suspected in previous tapered tri-coupler fabrications~\cite{KenchingtonGoldsmith2024}. The larger $W_L$~($=5$\,\si{\um}) simulation shows a familiar wavelength-dependent loss curve. This shows a larger loss of $>10$\% at the shortest wavelength that decreases to $<1$\% at the largest wavelength. It is included to highlight what may occur when optimising for a single wavelength (1800\,nm seen in Fig.\,\ref{fig:TaperedTriCoupler_Width1P8}).

\begin{figure}[ht]
    \centering
    \includegraphics[width=0.60\linewidth]{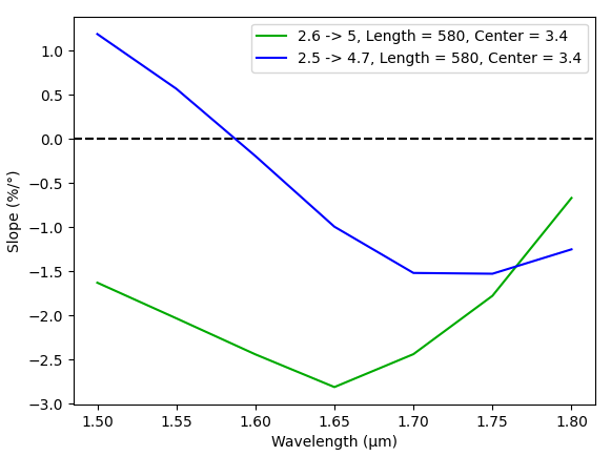}
    \caption{The slope of the starlight in the outer ports (the anti-null ports) at 180° phase difference. 0\,\%/° shows that a tri-coupler is not good for sensing.}
    \label{fig:TaperedTriCoupler_Sensing}
\end{figure}

Sensing in the nulling component of the interferometer is better described in Kenchington~Goldsmith~et~al.~\cite{KenchingtonGoldsmith2026}. The sensing slopes corresponding to parameter sets No. 1 and No. 3 from Table\,\ref{tab:WidthsSelection} are shown in Fig.\,\ref{fig:TaperedTriCoupler_Sensing}.

A slope of 0\,\%/° provides a region of degeneracy where it is impossible to tell the phase difference shift required to increase the null depth. In contrast, the higher slopes, whether negative or positive, provide greater control of the null depth during measurements. Without complete chromatic intensity/phase correction to provide the required input into the nulling component before the nuller, a single degeneracy in the waveband may not be an issue, or, at least, the most pressing issue for broadband nulling interferometry. Nevertheless, the figure demonstrates that parameter combinations exist for which no degeneracies occur. 

The curve corresponding to high loss at shorter wavelengths in Fig.\,\ref{fig:TaperedTriCoupler_ExThrough} shows a non-degenerate curve in Fig.\,\ref{fig:TaperedTriCoupler_Sensing}. This would simplify phase control across the full waveband, or at least provide a reliable estimate of the residual phase errors required to achieve a deep null. 

These results question whether there is a trade-off between maximising exoplanet throughput and maintaining sufficient phase-sensing sensitivity for active null control. Further experimental validation is required to determine whether this trade-off exists, or if other considerations have not been addressed when selecting the parameters for the nulling component of the nulling interferometer.

\subsection{Chromatically controlled achromatic intensity modulator}
\label{sec:Intensity_matching}


The wavelength-dependent phase response of the tapered directional coupler enables a new device concept: a chromatically controlled achromatic intensity modulator (CCAIM). Unlike a conventional Mach–Zehnder interferometer, the chromaticity in the thermo-optic phase shifter is deliberately compensated by a tapered directional coupler. Consider the standard matrix representation equation for an MZI:
\begin{equation}
    \begin{bmatrix} \text{Out}_1\\\text{Out}_2\end{bmatrix}= \begin{bmatrix} t_1 & ik_1\\ik_1 & t_1\end{bmatrix}\begin{bmatrix} \exp(i\theta) & 0\\0 & 1\end{bmatrix}\begin{bmatrix} t_2 & ik_2\\ik_2 & t_2\end{bmatrix}\begin{bmatrix} \text{In}_1\\\text{In}_2\end{bmatrix}
\label{eq:MZI}
\end{equation}
for two 2x2~couplers with cross-coupling ($k$) and transmissive ($t$) parameters such that $|k|^2+|t|^2=1$, and a thermo-optic phase shifter with an OPD shift ($\theta=2\pi \Delta L/\lambda$). 
The proposed device actively controls the optical power distribution between the two output ports. This can be used for intensity balancing in a nulling interferometer as an active optical attenuator, or as an optical switch, an active photometric channel, or as a mesh component for a coronagraph PIC~\cite{Sirbu2024}.

\begin{figure}[ht]
    \centering
    \includegraphics[width=0.16\linewidth,angle=-90]{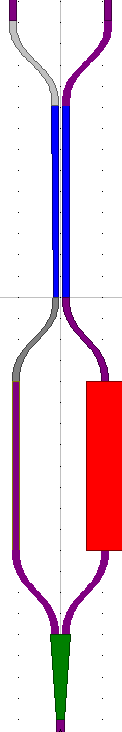}
    \caption{CAD of the chromatically controlled achromatic intensity modulator: a Mach-Zehnder interferometer using a Y-junction (Green), a tapered directional coupler (Blue), and a heater (Red) in one arm--above the waveguide offset in the Z-direction. The bent waveguides in the tapered directional coupler are colour-coded to indicate which have a width change (Grey and Light-grey) and which do not (Purple).}
    \label{fig:CAD_AIM}
\end{figure}

Figure\,\ref{fig:CAD_AIM} shows the Mach-Zehnder configuration for the CCAIM: A Y-junction producing two outputs with equal intensity and phase, a chromatic optical pathlength delay provided by a thermo-optic phase shifter, and a tapered directional coupler. In the simulations, the heater was represented by a metal strip whose refractive index was disabled to not perturb the optical mode calculations. 

\begin{figure}[ht]
    \centering
    \includegraphics[width=0.65\linewidth]{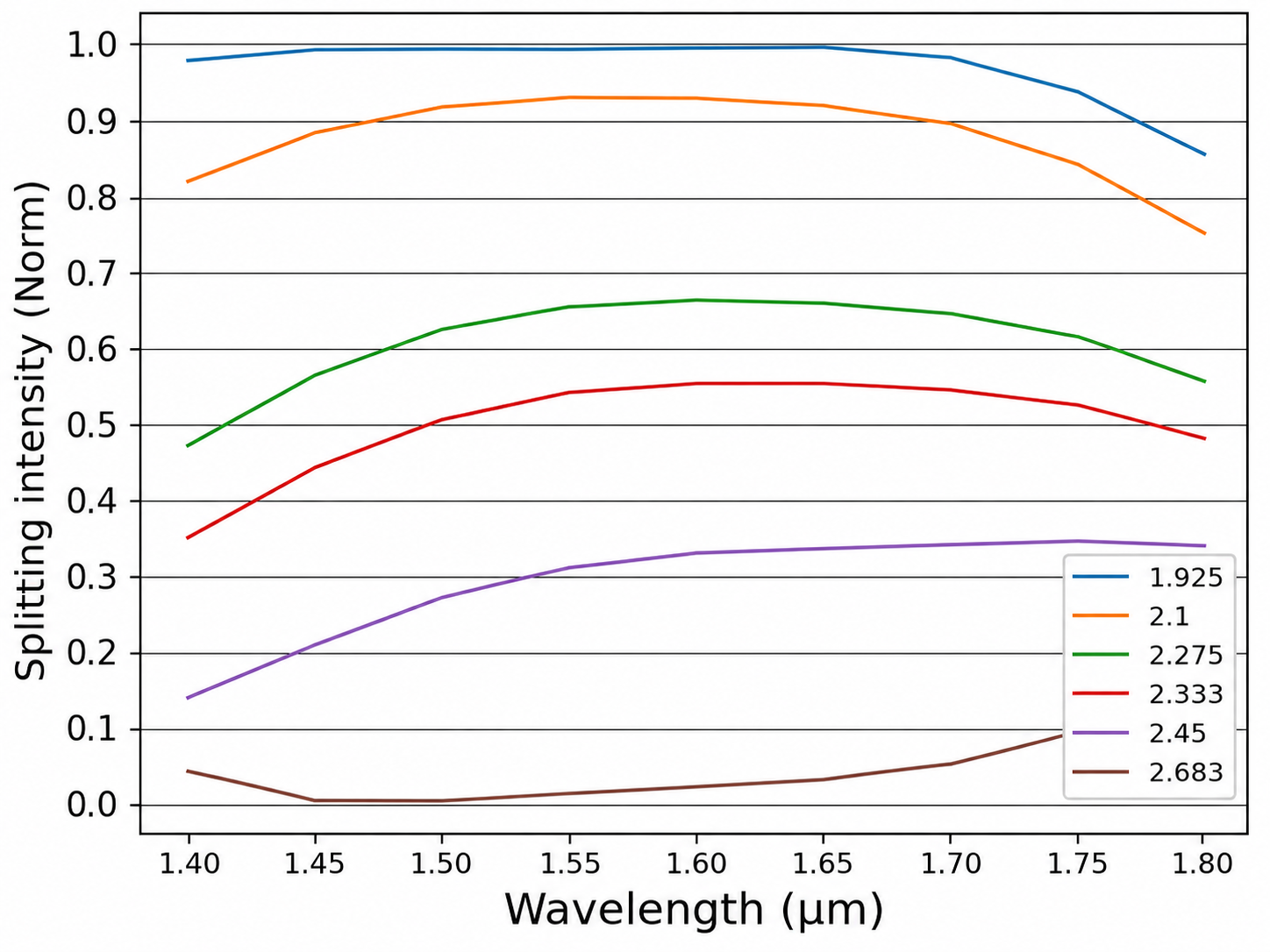}
    \caption{The splitting ratio in the CCAIM out of the left output port (normalised to the combination of output ports) for the labelled refractive index shift caused by the heater.}
    \label{fig:TaperedCombiner_IntensityShift}
\end{figure}

Figure\,\ref{fig:TaperedCombiner_IntensityShift} shows the simulated CCAIM labelled with the OPD from the heater. The OPD was adjusted to produce selected output splitting ratios ranging from 100:0 to 0:100.


The current proof-of-concept design achieves a nearly achromatic response only when all of the optical power is directed to a single output port: At an OPD of 1.92\,\si{\um}. An OPD of 2.62\,\si{\um} switches the light to the right output port, although it is the least achromatic curve. The OPDs in between split the light at varying levels of achromatism: the 35\% split providing a long (near-) achromatic response over 200\,nm. Thus, actively controlled achromatic intensity splitting can be achieved with a single heating source.

This configuration is unlike a tunable adiabatic couplers that use a similar tapered coupler with a heater interacting with the coupler itself~\cite{Liu2025}. Both approaches use a chromatic heater and chromatic coupler that work together to produce an achromatic splitting. The distinction lies in the operating principle. Liu~et~al.~\cite{Liu2025} use a single optical input and employ the heater to tune the coupler's splitting ratio directly, whereas the CCAIM combines two coherent inputs whose wavelength-dependent phase relationship is compensated by the tapered directional coupler.

From Eq.\,\eqref{eq:MZI}, it is not immediately obvious whether the achromatic response is governed primarily by the power splitting ratio of the tapered directional coupler or by the wavelength dependence of its phase response. The analysis presented below suggests that the latter is the dominant mechanism.

\begin{figure}[ht]
    \centering
    \begin{subfigure}{0.45\textwidth}
            \includegraphics[width=1.0\linewidth]{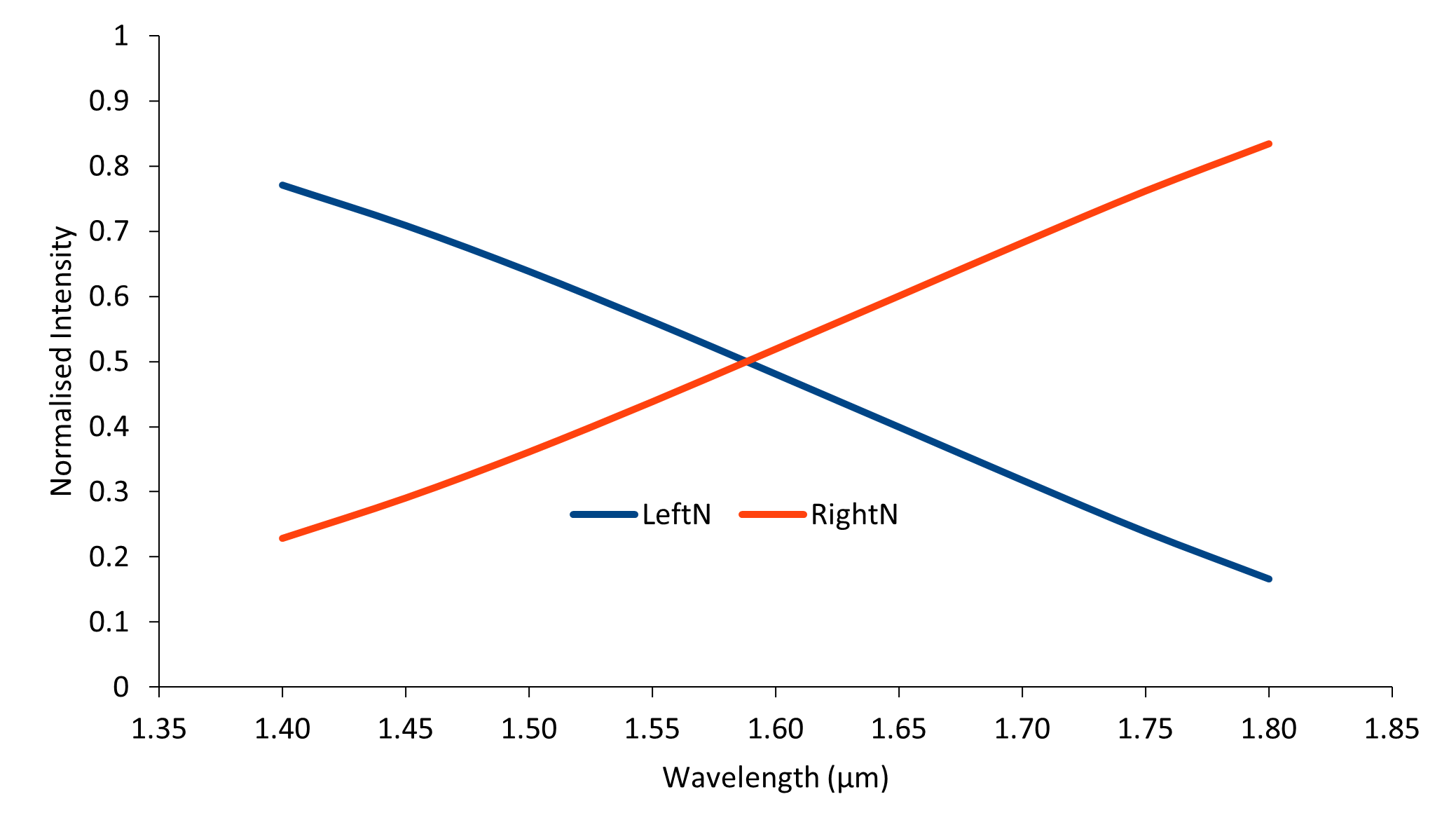}
        \caption{Standard directional coupler}
        \label{fig:DirectionalCoupler_Splittings}
    \end{subfigure}
    \begin{subfigure}{0.45\textwidth}
        \includegraphics[width=1.0\linewidth]{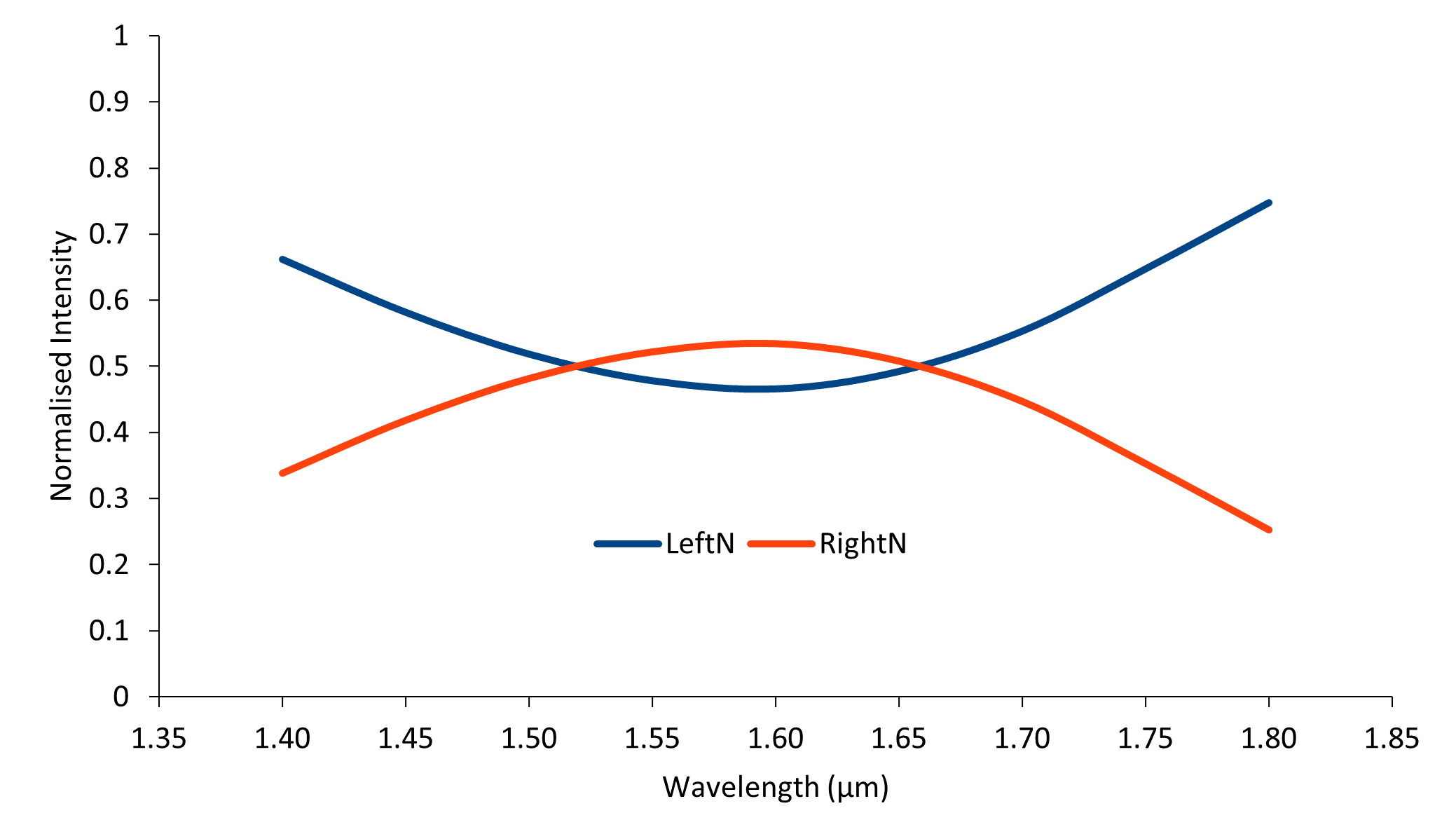}
        \caption{Tapered directional coupler}
        \label{fig:TaperedCoupler_Splittings}
    \end{subfigure}
    \caption{The power splitting ratio of (a) the standard directional coupler and (b) the tapered directional coupler.}
    \label{fig:Coupler_Splittings}
\end{figure}

Figure\,\ref{fig:Coupler_Splittings} shows the splitting ratio ($t$ and $k$) of the tapered beam combiner compared with a standard directional coupler.


The splitting shapes are distinct: the standard directional coupler has the expected cross and the tapered directional coupler exhibits a curved response, which looks like it could be flattened with a further parameter scan.

A natural question is whether a flatter wavelength response would broaden the achromatic bandwidth. To achieve this, the coupler would likely need to increase dramatically in length: In Sec.\,\ref{sec:TDC}, the tapered directional coupler took advantage of adiabatic tapering to increase the splitting ratio, but the interaction region in this proof-of-concept device is too short to achieve fully adiabatic behaviour. 

To obtain the broadest response possible within this compact geometry, the tapered arm used widths from 2.55 to 3.15\,\si{\um}, instead of matching the untapered waveguide's width in Sec.\,\ref{sec:TDC}, over a coupling region length of 910\,\si{\um}. These parameters have yet to be used to optimise the tapered directional coupler, beyond identifying this proof-of-concept design, and further optimisation will flatten this curve and potentially improve the CCAIM. However, the wavelength dependence of the power splitting alone does not determine the CCAIM performance. Equally important is the wavelength dependence of the coupler phase.

To illustrate this, Eq.\,\eqref{eq:MZI} is recast as
\begin{equation}
\begin{bmatrix} \text{Out}_1\\\text{Out}_2\end{bmatrix}= \begin{bmatrix} t\exp(i\phi) & k\exp(i\gamma)\\k \exp(i\gamma) & t\exp(i\phi)\end{bmatrix}\begin{bmatrix} \exp(i\theta) & 0\\0 & 1\end{bmatrix}
    \label{eq:CCAIM}
\end{equation}
where $\phi$ and $\gamma$ are the tapered directional coupler's phase components that may or may not be related. 

The transmission and coupling coefficients ($t$ and $k$) are obtained from single-input BeamPROP simulations in Fig.\,\ref{fig:Coupler_Splittings}, and rather than explicitly calculating $\phi$ and $\gamma$, two equal-amplitude inputs with varying phase difference were injected into the coupler to determine the input phase-to-power splitting relationship.

\begin{figure}
    \centering
    \begin{subfigure}{0.45\textwidth}
            \includegraphics[width=1.0\linewidth]{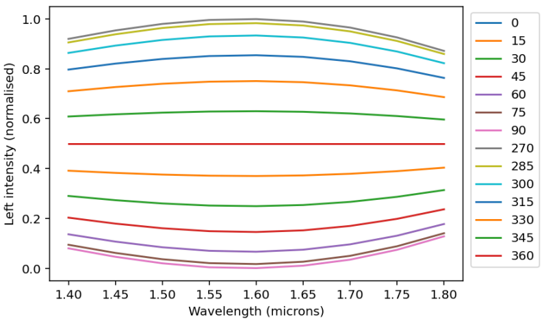}
        \caption{Standard directional coupler}
        \label{fig:TypicalDirectionalCoupler_Phase}    
    \end{subfigure}
    \begin{subfigure}{0.45\textwidth}
            \includegraphics[width=1.0\linewidth]{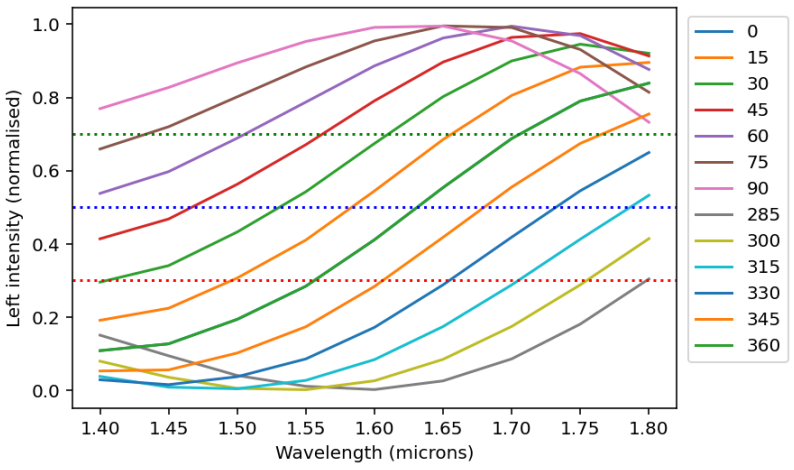}
        \caption{Tapered directional coupler}
        \label{fig:TaperedDirectionalCoupler_Phase}
    \end{subfigure}
    \caption{The output splitting intensity for (a) the standard directional coupler and (b) the tapered directional coupler when injecting equal electric fields at the labelled phases.}
    \label{fig:Coupler_Phase}
\end{figure}

Figure\,\ref{fig:Coupler_Phase} shows the optical power emerging from the left output port when two equal-intensity electric fields are input into a tapered directional coupler with different phase differences, with a standard directional coupler for comparison.

As expected from the conventional coupling matrix, in Eq.\,\eqref{eq:MZI}, the standard directional coupler couples light equally with a 0° (=360°) phase shift, and switching light to one arm requires either a 90 or 270° phase difference. Away from the central wavelength (1.6\,\si{\um}), the wavelength dependence of the output splitting becomes increasingly pronounced as the condition $t= k= 0.5$ is no longer satisfied. 

In contrast, the tapered directional coupler has a continuous wavelength-dependent response for every phase difference. Thus, to get the same equal-power beam combination achromatically, one would require a chromatic phase input into the coupler.

\begin{figure}[ht]
    \centering
    \includegraphics[width=0.6\linewidth]{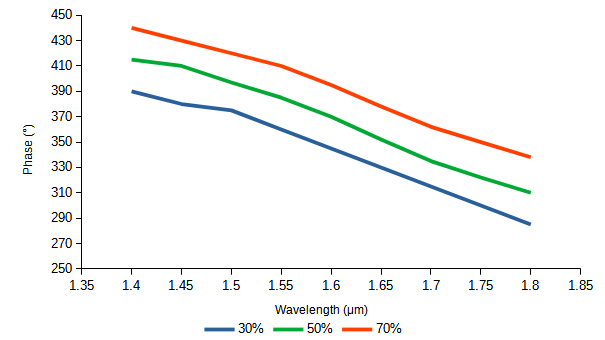}
    \caption{The unwrapped phases from Fig.\,\ref{fig:TaperedDirectionalCoupler_Phase} required to achieve a 30, 50, and 70\% splitting ratio.}
    \label{fig:PhaseRequirments}
\end{figure}

The dotted curves in Fig.\,\ref{fig:TaperedDirectionalCoupler_Phase} are included to guide the eye. The intersection of the dotted guide curves with the solid simulation curves yields the required input phase at each wavelength for achromatic splitting ratios of 30, 50, and 70\%. The cross-section of the dotted lines and the solid lines is shown in Fig.\,\ref{fig:PhaseRequirments}. 

Therefore, designing a thermo-optic phase shifter that reproduces the phase profiles in Fig.\,\ref{fig:PhaseRequirments} is sufficient to generate achromatic splitting. Consequently, optimising the coupler geometry to tailor this phase relationship is likely to be more important than simply flattening the wavelength dependence of the power splitting. However, the two objectives are expected to be related.



\begin{figure}[ht]
    \centering
    \includegraphics[width=0.7\linewidth]{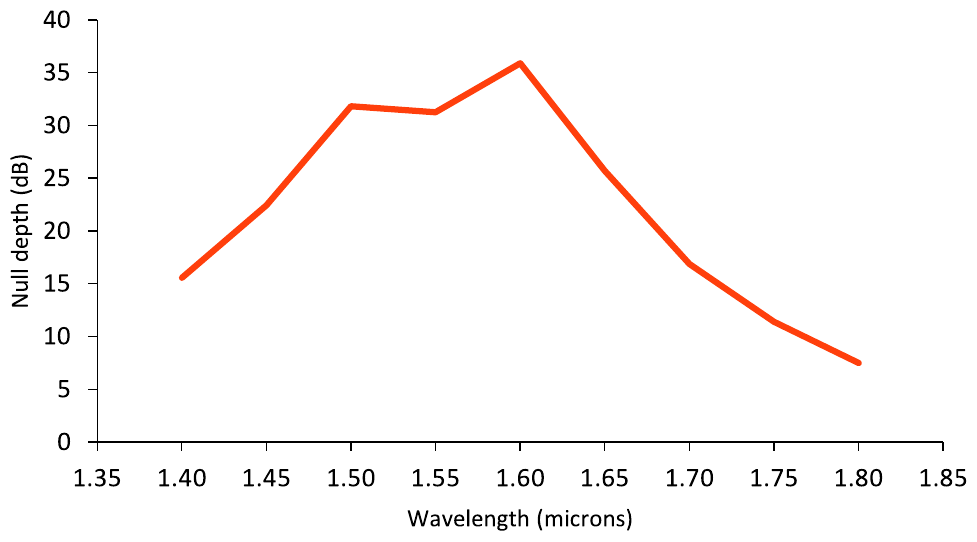}
    \caption{Calculated null depth for the CCAIM with an OPD of 1.92\,\si{\um}, obtained from the simulations in Fig.\,\ref{fig:TaperedCombiner_IntensityShift}. The null depth compares the power emerging from the left output port for 0° phase difference (exoplanet) and 180° out-of-phase (starlight) inputs.}
    \label{fig:null}
\end{figure}

The CCAIM has been demonstrated here as an intensity switch. It may also function as a standalone nulling interferometer. Figure\,\ref{fig:null} demonstrates a simulated extinction ratio exceeding 30\,dB across a 200\,nm waveband using an OPD of 1.92\,\si{\um}. This corresponds to the wavelength range over which the tapered directional coupler maintains its approximately 40:60 splitting ratio. Although this application requires further investigation and is beyond the scope of these proceedings, it demonstrates that the tapered directional coupler may also function as the beam-combining element of a nulling interferometer.

Further optimisation is required to determine whether the CCAIM can achieve an achromatic response across the full range of programmable splitting ratios, or whether alternative MZI architectures based on conventional components can provide improved performance. The approximately 150\,nm achromatic bandwidth obtained for programmable splitting ratios between 100\% and 35\% is comparable to that reported for an adiabatic coupler~\cite{Liu2025}, although it is achieved with a substantially larger footprint.


\section{Future directions: a mid-infrared chalcogenide platform}
\label{sec:MIR}

The broadband components demonstrated in the previous sections enable the foundation for a future mid-infrared photonic platform. This section outlines a proposed roadmap towards an integrated ChG nulling interferometer for astronomical applications.

The MIR is a key wavelength regime for future astronomical interferometry that could expand the scope of several proposed space missions, Habitable Worlds Observatory~\cite{Feinberg2024} and LIFE~\cite{Quanz2022}, and ground-based telescopes like the VLTI. Existing MIR astronomical instruments include MATISSE~\cite{Lopez2022} and NOTT~\cite{Defrre2024} at the VLTI, the latter of which uses a photonic chip in the L-band~\cite{Sanny2026} for nulling interferometry.  

The MIR is particularly attractive for nulling interferometry because the longer wavelengths relax diffraction constraints, improving access to planets at small angular separations for detection and analysis for signs of life on these exoplanets~\cite{Selsis2008}, assuming enough of the starlight can be suppressed.

\begin{figure}[ht]
    \centering
    \includegraphics[width=0.9\linewidth]{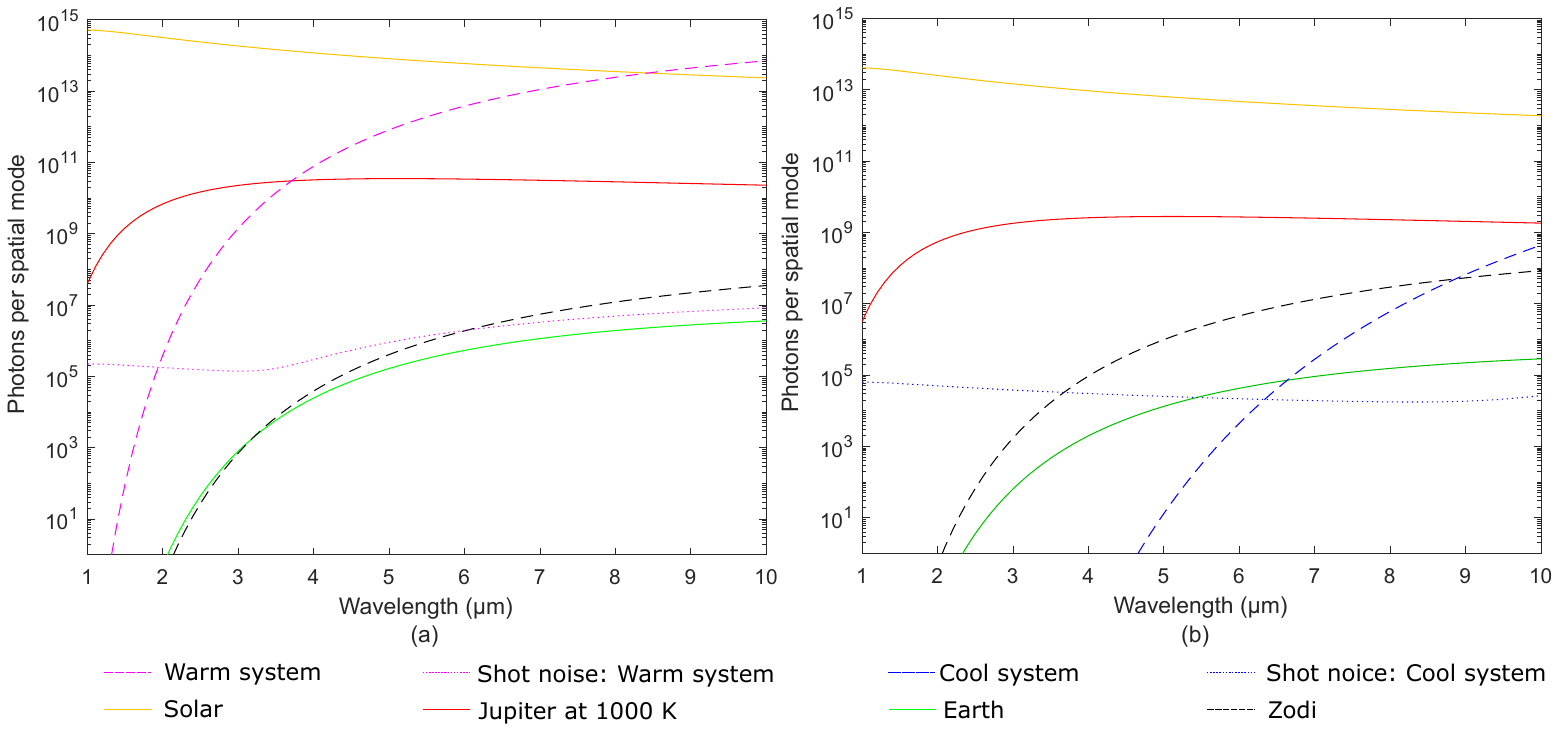}
    \caption{Obtainable photons in (a) a warm system and (b) a cool system. A cold system uses a 2\,m telescope and 24\,h of total integration time. The warm system used an 11\,m telescope and 10\,h of integration time. Both assumed a 35\% efficiency, 10\% emissivity, and a 16\% bandwidth. The Earth, 1000\,K Jupiter, and Solar (in green, red, and yellow solid lines) analogue was simulated 10\,pc away. The system light (in dashed lines) from the telescope system at 280\,K (magenta) and 80\,K (blue) are shown alongside the Zodiacal light (black). The shot noise limit (in dotted lines) represents the limitations of the respective systems with a 40\,dB null over the star.}
    \label{fig:Exoplant_detection}
\end{figure}

Figure\,\ref{fig:Exoplant_detection} compares a warm system, on Earth, with a cold system, at Lagrange Point Two (L2), for detecting exoplanet photons. It includes an equivalent Sun and Earth system with a Jupiter-type exoplanet at 1000\,K, all 10\,pc away. The background light is also included for the warm and cold systems, and Zodical light from our solar system, with the resulting shot noise limit. 

The shot noise is the combination of photons from the system, zodi, and starlight; the latter was simulated to have a 40\,dB ($10^4$) suppression over the star. Signals exceeding the shot-noise limit are, in principle, detectable. It shows that an Earth-like exoplanet 10\,pc away cannot be detected for an 11\,m telescope, whereas for a 2\,m space telescope, above a 5\,\si{\um} wavelength, it is. Hot Jupiter-type exoplanets, on the other hand, can be detected at all wavelengths.


These simulations indicate that the N-band provides a viable pathway to Earth-like exoplanet detection when measured from space. Using a deeper null depth may not assist in this since it is the warm system light causing the limitation, but that doesn't mean other exoplanets cannot be found using ground-based telescopes -- Jupiter-style exoplanets are in the detection range. Space-based observatories therefore provide the most promising route for Earth-like exoplanet detection. These require lightweight components to reduce cost, and photonics offers complex circuitry on a light monolithic PIC. 

The NOTT PIC uses a Bracewell nulling architecture on a 3-dimensional, ChG platform. This is the inspiration for a four-telescope two-dimensional interferometer, first creating an L-band NOTT equivalent before moving to the N-band. The NOTT PIC uses a low index contrast type architecture, directly written into a ChG block using ultra-fast laser inscription~\cite{Gretzinger2019a},  with an even lower index contrast than the SiO\textsubscript{2} platform discussed in Sec.\,\ref{sec:Caltech}, which makes it easier for mode matching but not for compact circuits, leaving little optimisation from the current system. 

Higher index contrasts can be tailored for different wavelength regimes and applications by using different ChGs. Sulphur (S-), selenium (Se-), or tellurium (Te-) are the chalcogen elements in the glass, and each has, on average, a higher base refractive index than the last. A combination of S-ChG and Se-ChG should provide a high index contrast equivalent to that in Sec.\,\ref{sec:PLANETES}, for example. A high index contrast Bracewell nuller was designed using a lithographic germanium, arsenic, and selenium on germanium, arsenic, and sulphide, step-index, two-dimensional platform~\cite{KenchingtonGoldsmith2016}. The MMIs fabricated for this platform exhibited excessive loss for this application, and the injection loss was never overcome. 

This proposed platform instead adopts tapered directional couplers and tapered tri-couplers, which are designed to minimise excess loss, for splitting and recombination. The tapered tri-coupler is the obvious candidate as a nulling beam combiner, but these need to be tested and fabricated in the lab to understand whether there are fabrication limitations to the null depth. Tapered directional couplers should also be tested as ABCD-couplers for a pairwise interferometer, the equivalent of the PLANETES project (in Sec.\,\ref{sec:PLANETES}), but for the L-band. The beam combiner for the MATEESE instrument is the bulk optical analogue. 

The broadband tapered tri-couplers and directional couplers presented in this work provide the key enabling components for this future platform, replacing conventional MMIs, directional couplers, and Y-junctions with devices that offer lower excess loss and improved broadband performance.

The larger issue to overcome is the insertion loss. For this, there are three pathways of increasing complexity:
\begin{itemize}
    \item Adiabatic spot-size converters
    \item Waveguide-array mode expanders
    \item Integrated or bonded microlens arrays
\end{itemize}

Two complementary taper strategies will be investigated: a large multimode waveguide that is tapered down to the single-mode width adiabatically to retain as much light as possible, or a small waveguide that squeezes the mode confinement such that the fundamental mode matches the insertion mode field. Both attempts have their pros and cons and will be investigated to see which is more effective in practice. A waveguide array expands the total mode size, effectively creating a large, weakly guided waveguide that requires multiple layers of lithography. After which, all but the main waveguide are tapered away such that the light is squeezed into the single-mode waveguide, mitigating loss where possible. A lens array, or mounting microlenses to a photonic chip, is a well-established method in photonics for injecting light into a chip. It is the most practical of the three pathways but requires precise alignment and novel methods of attachment since standard polymers usually used for attaching microlenses absorb in the MIR.  

For the N-band, these materials will not be compatible due to the S-absorption features in S-ChG past a 7\,\si{\um} waveguide. Alternative ChG compositions identified by Kenchington~Goldsmith~et~al.~\cite{KenchingtonGoldsmith2018}  indicate low-loss Se-ChG that can extend from the L-band into the N-band. These materials will be investigated first before attempting ChG with Tellurium (Te-ChG). Te-ChG generally exhibits higher propagation losses~\cite{Hilton1970}, making the Se-ChG the optimal material for the N-band.

Together, these developments define a roadmap towards integrated L-band and ultimately N-band photonic beam combiners for future ground-based and space-based nulling interferometers.

\section{Conclusion}
\label{sec:Conclusion}

This work demonstrates that broadband photonic splitters and couplers can be realised for both low- and high-confinement waveguide platforms, represented here by SiO\textsubscript{2} and Si\textsubscript{3}N\textsubscript{2} technologies operating in the astronomical H- and J-bands, respectively. The tapered tri-coupler has been demonstrated as a broadband splitter that inherently provides achromatic starlight suppression when used as a nulling beam combiner while maintaining greater than 95\% exoplanet throughput.

For the Si\textsubscript{3}N\textsubscript{2} platform, optimised tapered tri-couplers achieved less than 1\% loss across the full J-band. The equivalent SiO\textsubscript{2} design achieved less than 2.2\% exoplanet throughput loss across the H-band while preserving broadband nulling performance.

Optimised tapered directional couplers provided broadband 50:50 and 40:60 splitting suitable for pairwise beam combination. Achromatic 50:50 operation required a longer coupling region than the 40:60 design, illustrating the trade-off between compactness and broadband performance.

This work also introduces the chromatically controlled achromatic intensity modulator. The preliminary results show promise. Using a single thermo-optic heater, a tapered directional coupler, and a Y-junction, a programmable attenuation between 100 and 35\% was simulated to be achromatic over 300 to 150\,nm, respectively. Further loss compensation becomes increasingly chromatic, and more work is required to determine whether these curves can be flattened by improving the tapered directional coupler's splitting ratio.

The mid-infrared regime is the next step in this work. Extending these low-loss broadband components to the mid-infrared will provide a foundation for multi-telescope photonic interferometers operating in the L- and N-bands, supporting future astronomical facilities such as HWO and LIFE.

Together, these developments illustrate the growing maturity of broadband photonic components for astronomical interferometry, including nulling interferometry. Continued progress in materials, fabrication precision, and device design will be essential for enabling fully integrated, low-loss, and highly scalable photonic interferometers across the near- and mid-infrared spectrum.

As astronomical PICs continue to evolve for deployment on large ground-based observatories and future space telescopes, their photonic circuitry must become increasingly sophisticated and scalable, particularly for interferometric architectures that require beam combination from many telescope apertures or segments.

The broadband passive and active photonic components presented here provide a pathway towards scalable, fully integrated interferometric instruments spanning the near- and mid-infrared, bringing astronomical photonics closer to the requirements of the next generation of ground- and space-based observatories.


\section* {Acknowledgments}
The SiO\textsubscript{2} work was supported by the National Aeronautics and Space Administration~(NASA) under grant number 80NSSC24K1559, awarded through the Astrophysics Research and Analysis~(APRA) program. This work was supported by the National Science Foundation under Grant numbers 2308360 and 2308361. 

This Si\textsubscript{3}N\textsubscript{2} work was supported by the ERC Advanced Grant 101142746 and the funding from the project PHOTONICS, which was financed by the ANR program PEPR Origins (ANR-22-EXOR-0005).



\bibliography{report,Extra}   

\bibliographystyle{spiebib} 

\end{document}